\documentclass{article}

\usepackage{arxiv}

\usepackage[utf8]{inputenc} 
\usepackage[T1]{fontenc}    
\usepackage[colorlinks=true, allcolors=blue]{hyperref}       
\usepackage{url}            
\usepackage{booktabs}       
\usepackage{amsfonts}       
\usepackage{nicefrac}       
\usepackage{microtype}      
\usepackage{lipsum}		

\usepackage{graphicx}

\title{Serverless on FHIR: Deploying machine learning models for healthcare on the cloud}

\author{
  Bell Raj Eapen\\
  Information Systems\\
  McMaster University\\
  Hamilton, ON, Canada \\
  \texttt{eapenbp@mcmaster.ca} \\
    \And
 Kamran Sartipi \\
 Department of Computer Science \\
East Carolina University \\
Greenville, NC, USA \\
 \texttt{sartipik16@ecu.edu} \\
    \And
 Norm Archer \\
 Information Systems\\
 McMaster University\\
 Hamilton, ON, Canada \\
 \texttt{archer@mcmaster.ca} \\
}

\begin{document}
\maketitle

\begin{abstract}

Machine Learning (ML) plays a vital role in implementing digital health. The advances in hardware and the democratization of software tools have revolutionized machine learning. However, the deployment of ML models --- the mathematical representation of the task to be performed --- for effective and efficient clinical decision support at the point of care is still a challenge. ML models undergo constant improvement of their accuracy and predictive power with a high turnover rate. Updating models consumed by downstream health information systems is essential for patient safety. We introduce a functional taxonomy and a four-tier architecture for cloud-based model deployment for digital health. The four tiers are containerized microservices for maintainability, serverless architecture for scalability, function as a service for portability and FHIR schema for discoverability. We call this architecture \emph{Serverless on FHIR} and propose this as a standard to deploy digital health applications that can be consumed by downstream systems such as EMRs and visualization tools.

\end{abstract}


\section{Introduction}

Artificial Intelligence (AI) and Machine Learning (ML) are transforming every aspect of healthcare from diagnostics to therapeutics \cite{yu2018artificial}. The capabilities of many of these applications are maturing at a fast pace. There is a growing interest in building ML models --- the mathematical representation of the task to be performed --- that have a positive impact on patient care. With the democratization of ML techniques and the increasing availability and accessibility of high-performance computing, increasingly sophisticated ML models in healthcare are being built. However, making the application available at the point of patient care --- operationalizing the ML model --- in a timely and efficient manner is still a challenge as most models have a short life span requiring constant revisions. Many of the sophisticated and expensive ML models end up as theoretical artifacts in academic publications with minimal practical utility \cite{chen2019develop}.

Artificial Intelligence (AI) --- as used here is synonymous with operationalized ML models --- may currently be at the \emph{peak of expectations} of the Gartner hype cycle \cite{o2008gartner}. Even after the hype wanes off, AI will have applications in several healthcare domains. AI has matured in sectors such as business analytics and banking. Some of the successful tools and techniques from these domains can be applied to improve the application of AI in healthcare \cite{ben2019impact}.


ML can be used in fundamental research (e.g. bioinformatics and computational biology) \cite{tang2019recent}, clinical practice (e.g. clinical decision support systems)\cite{shen2017deep} or healthcare delivery (e.g. prediction of healthcare consumption) \cite{kwakernaak2020using}. In this article, we will be focusing on AI in clinical practice. Functionally, AI in clinical practice can be used for diagnosis (identifying the presence of a named disease entity) \cite{sboner2003multiple}, prognosis (predicting future risk) \cite{kourou2015machine}, therapeutics (treatment) \cite{weng2002protein}, and public health \cite{eapen2020public} to name a few. AI has applications in several specialties such as Radiology (image analytics) \cite{wang2012machine}, Dermatology (skin cancer classifiers) \cite{sboner2003multiple}, Cardiology (interpretation of echocardiography) \cite{roopa2017survey} and Neurology (analyzing electroencephalograms) \cite{muller2008machine}. Developing AI for digital health involves selecting an appropriate problem, identifying and curating the appropriate datasets, developing and evaluating models and finally assessing the clinical impact \cite{chen2019develop}. Problems, datasets and evidence change with time, causing old models to lose relevance. It is vital to update and maintain models used in healthcare to ensure patient safety \cite{obermeyer2016predicting}.

The rest of the article is structured as follows: First, we propose a workflow-centric taxonomy to support implementation. This modified taxonomy will serve model deployments better than the functional taxonomies as described above. We then extend techniques from cloud-based business intelligence to solve common problems such as maintainability, scalability, portability, and discoverability of ML model deployments. Next, we offer some pragmatic suggestions for model deployment, and model management, by embedding models into clinical workflow. FHIR is emerging as a popular standard for healthcare interoperability. We propose FHIR schema for the integration of ML application programming interfaces (API) into existing health information systems. Finally, we demonstrate the process with an example.

\section{Taxonomy of ML models}

The various ML algorithms such as regression, decision trees and neural networks, typically build a mathematical model of the task to be performed after the training phase \cite{kaur2006empirical}. Most platforms used for machine learning will allow the exporting of this representation in a proprietary or open format \cite{geron2019hands}. These exported models contain the weights required for the prediction task along with other metadata generated during the building phase.

Models in healthcare can be varied based on factors such as the ML method used to create them. From a clinical perspective, ML models are classified according to the domain in which they are used such as diagnostic models (that ascertain the presence of a known disease) \cite{kim2017development}, prognostic models (that estimate the future outcome for a given patient) \cite{betancur2018prognostic}, therapeutic models (that predict the best treatment given the set of probable diagnosis and patient-related constraints) \cite{toussi2009using} and public health models (that predict group outcome such as an outbreak of an epidemic) \cite{nair2018applying}. This clinical classification guides ML model builders in healthcare to isolate the problem and propose the best prediction task. 

This classification is inadequate for the technical team concerned with functions such as model deployment, model management, designing APIs and ultimately making the deployed model to be available to clinicians at the point of patient care. We propose a technical taxonomy for ML models in healthcare drawing from our information system experience along with typical examples. 

\subsection{Explanatory models}

Explanatory models attempt to ‘explain’ a phenomenon by specifying the relationships among a setup of constructs given a boundary condition. The constructs are generally concepts that cannot be measured and are calculated from a set of variables that are directly measured. Explanatory models are widely used in behavioural sciences and psychology \cite{yarkoni2017choosing}. Techniques for dimension reduction such as principal component analysis (PCA) are commonly used to build explanatory models.

Models encompassing tree or rule-based logic such as decision trees and random forests are explanatory. Decision trees are particularly important in healthcare owing to their simplicity and ease of understanding. Linear and logistic regression models are also explanatory in nature providing insights on the relative importance of various factors in deciding the final outcome or dependent variable.

Explanatory models can be built using various statistical and machine learning packages and exported to proprietary formats. Irrespective of the nature of the exported artifact, most explanatory models take a set of inputs, process it in a manner that can be explained and give one or more outputs.

The decision tree model for classifying cardiac arrhythmias based on a feature set is an example of an explanatory model \cite{sahoo2020automatic}.  Cardiac arrhythmias --- irregular heartbeat pattern --- can be life-threatening. Manual arrhythmia classification from ECG signals is hard because of its non-stationary nature. A proposed decision tree model in a published example \cite{sahoo2020automatic} for classifying six types of arrhythmias based on 15 features had an accuracy of 99\%. The decision algorithm based on the 15 features is explainable. 

Explainable models typically have a longer life span and are easier to deploy compared to predictive models described below.

\subsection{Predictive models}

Predictive models prioritize predictive power over explanatory power \cite{shmueli2011predictive}. Predictive models require different modelling approaches so that they are fine-tuned for maximum predictive power.  Most probabilistic models such as neural networks and distance-based models such as clustering fall under this category. Predictive models are popular in healthcare because of their practical utility and because some are partially interpretable \cite{choi2016retain}.

Popular skin cancer screening apps make use of such predictive models \cite{freeman2020algorithm}. For example, malignant melanoma is aggressive skin cancer, but it is difficult to differentiate it from a benign nevus (normal mole), especially in the early stages. Deep learning models, especially the convolutional neural networks (CNN)  based on images of histopathology slides can be as accurate as trained clinicians in identifying melanoma \cite{hekler2019deep}. However, the CNN models trained on labelled images of melanoma and nevi are black boxes that cannot be explained. 

More accurate models are becoming available on a regular basis through the use of more data and better algorithms. Hence, predictive models need to be updated regularly to ensure patient safety \cite{varshney2017safety}.

\subsection{Environment models}

Reinforcement learning (RL) is an emerging paradigm in machine learning concerned with software agents learning to take actions in a particular environment in order to maximize some reward \cite{sutton2011reinforcement}. An agent learns the best actions for a given situation (policy) by dynamically interacting with the environment. To accomplish this, the agent needs a representation of the environment that provides information on its changing states based on the agent’s actions and the associated reward. In RL parlance, an environment is called a model only when it is complete (model-based RL). Here, we apply the term model to any representation of the environment that can be used for RL.

Sepsis is a potentially lethal infection that requires the optimum combination of interventions to improve the patient’s chances of survival \cite{saria2018individualized}. These interventions have to be administered at the right time, depending upon various patient characteristics where early or late administration of an intervention may be disastrous. For example, fluid administration increases tissue perfusion, but at the same time increases the chances of edema. Existing evidence on the probabilities of different outcomes for patient attributes can be encapsulated as an environment model. RL agents can learn from such models, trying to optimize some notion of reward which is patient survival in the case of sepsis.

Environment models have low life spans and need to be updated more frequently than predictive models as more data become available.

\subsection{Interaction models}

Interaction models are used in the context of conversational agents (chatbots) that understand user utterances and provide relevant responses. Chatbots are gaining popularity in healthcare, mostly for providing behavioural interventions and triage  \cite{csaky2019deep}. Interaction models may maintain state during the course of a conversation with a client (‘slots’ in the chatbot parlance), but these states are generally temporary. Interaction models classify the client’s ‘intent’ and produce an output for the inferred intent based on the filled slots – i.e. the background information available to the agent  \cite{roca2020microservice}. Interaction models can be shared across agents with similar functions. The concept of transfer learning may be applicable in the context of conversational agents whereby specialized agents are designed by extending generic interaction models that may be applicable in multiple related situations.

The models generated by ML experts need to be deployed in an efficient and sustainable manner to be useful to the clinician, who ultimately uses these models for patient care  \cite{wallace2012deploying}. We have not included in our taxonomy self-learning models that continuously evolve and learn from data streams. Agents in reinforcement learning are typical examples of such models. Models that we describe typically do not maintain their state across different sessions. For example, interaction models in chatbots keep state during a session but do not usually save the session for reuse and hence do not require data persistence.

In the following discussion we describe deployment guidelines for models in healthcare that do not need access to persistent storage for saving state. We have not included self-learning and session-preserving models in our taxonomy as the proposed guidelines do not apply to models that save their state.

\section{Deployment Guidelines}

The utility and effectiveness of ML models in healthcare depend on their availability at the point of care \cite{wallace2012deploying}; their maintainability as new models emerge and scalability as the demand waxes and wanes. The adoption of suitable standards enables the sharing of models. DevOps — the process of providing continuous delivery of modified software artifacts — in the context of ML models is becoming increasingly important \cite{amershi2019software}. We posit maintainability, sustainability, portability and discoverability as the four rungs of a pyramid that are vital in the effective and efficient DevOps of any ML model in healthcare. 

\begin{figure}[ht]
\centering
\includegraphics[width=0.9\textwidth]{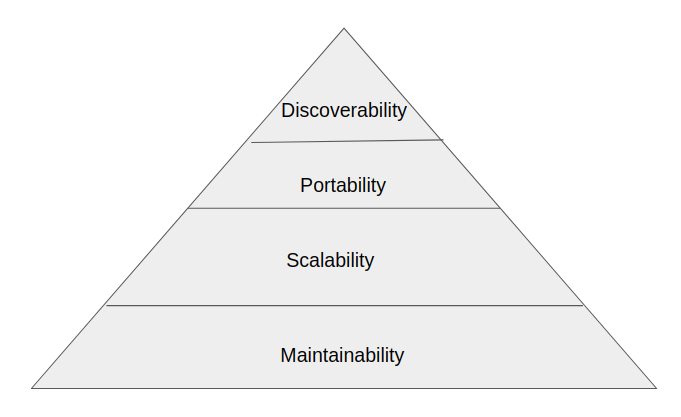}
\caption{Serverless on FHIR pyramid}
\label{fig:pyramid}
\end{figure}

\subsection{Maintainability}

Maintainability has several dimensions in software engineering, but in the context of ML models, it represents the ease with which modifications can be made to the artifact to correct faults or to improve performance after deployment  \cite{malhotra2016software}. Continuous improvement, one of the aspects of maintainability,  is especially crucial for ML models in healthcare because of patient safety implications. New and better ML models emerge all the time in healthcare and some can outperform human decision-makers. Any delay in synchronizing deployed models to the most sensitive and specific ML model available can potentially affect human lives. 

Software maintainability research has identified microservices architecture as a significant factor in its ease of deployment \cite{dragoni2017microservices}. Microservices architecture splits up monolithic applications into small, independent, distributed and strongly decoupled parts  \cite{dragoni2017microservices}. Microservices, owing to their modular nature is easy to be maintained. The availability of containerization tools such as docker \cite{merkel2014docker} facilitates the adoption of a microservices architecture. Containerization is the process of packaging an application and all its dependencies and environment in a manner that facilitates deployment on various computing environments with minimal changes \cite{kang2016container}. Containerization is the cornerstone of DevOps — providing continuous software delivery with high quality \cite{ebert2016devops}.

Self-adaptive orchestrating platforms such as Kubernetes \cite{bernstein2014containers} and Docker Swarm \cite{modak2018techniques} can efficiently manage containerized microservices. The use of small, stateless, containerized, microservices built around a single clinical problem tends to promote maintainability.

\subsection{Scalability}

The scalability of model deployments is important in healthcare applications. As its use increases, the system should be able to deal with the increased load without increases in latency and impact on patient care. Similarly, as models change with changing evidence, their use may decline, and the system should downscale accordingly to conserve resources \cite{weyuker2002metric}. Emerging serverless architecture in cloud computing \cite{baldini2017serverless} --- where the cloud service provider manages the backend --- that is capable of scaling from 0 to infinity. However, Cold start — delay in the availability of a function after invocation — is a problem in serverless operations \cite{manner2018cold}, especially after a service is completely shut down (scaled to 0) due to low usage. A serverless function is a small, reusable chunk of code (Function as a Service \cite{lynn2017preliminary}) that is stateless and can be easily deployed on a suitable infrastructure. The serverless architecture supports both upscaling and downscaling of model deployments according to actual use. Serverless architecture is well suited for healthcare model deployments that scale with usage. 

\subsection{Portability}

Portability is used here to represent the ease with which a deployment in one cloud provider platform can be moved to another provider. However, vendor lock-in is a problem with cloud providers that reduces flexibility and increases software cost. There have been several attempts by vendors and other companies to promote standards that will ensure portability. One such popular standard stipulates deployment in 64-bit Linux containers not exceeding 2 GB of memory, listening to and responding on port 8080 \cite{kaewkasi2018docker}. In this solution, computation is stateless and scoped to a request. Such standardized containers can be easily deployed on a serverless framework. OpenFaaS provides a set of tools to facilitate the management of such standardized containers \cite{ellis2019openfaas}; and Google has followed suit with the Google Cloud Run \cite{gcr2019serverless}. Containers built according to a standard can reduce vendor lock-in for ML workflows in healthcare.

\subsection{Discoverability}

Downstream systems such as Electronic Medical Record systems (EMRs) need to communicate with models deployed in the cloud. Ideally, downstream systems should be able to discover the input and output parameters of a deployed model, so that both can exchange information without any configuration. This would allow ML workflows to be used within the context of EMRs and other health information systems in a ‘plug-and-play’ manner.

Fast Health Interoperability Resource (FHIR) has emerged as the de-facto standard for the exchange of information between health information systems \cite{eapen2019drishti}. The draft standard for trial use (DSTU) stage is over, and the current version (R4) is considered stable. FHIR’s popularity is due to the use of existing web standards such as the RESTful (Representational state transfer) interfaces and the XML or JSON (JavaScript Object Notation) format for data exchange. FHIR based systems exchange  \emph{Resources} that represent a semantic entity represented by a defined set of elements and metadata that supports 80\% of the use cases \cite{bender2013hl7}. For example, name, gender and date of birth are attributes for the  \emph{Patient} resource. The  \emph{Endpoint} resource describes the technical specifications of an information system that can connect and exchange other FHIR resources. It consists of elements such as address or the uniform resource locator (URL), header and the type of the  exchanged payload.  \emph{Observations} are measurements and assertions made about a patient having attributes such as the value and the reference range.  \emph{CarePlan} represents the plan of how one or more clinicians intend to deliver care to a particular patient with attributes such as activity and detail. A FHIR  \emph{Bundle} is a collection of related resources that can be exchanged as a single unit.

FHIR EndPoint resource is ideally suited for exposing the service details in a declarative manner for discoverability by consuming systems \cite{fhir2019endpoint}. Any GET request on the  \emph{Serverless on FHIR} should produce a FHIR EndPoint resource that describes the expected input and output. The input to the  \emph{Serverless on FHIR} should be a POST request. Input and output should produce a FHIR bundle as JSON that typically consists of a Patient resource, Observations, CarePlans and Subscription resources. The uniform format of the input and output allows the chaining of ML models as a pipeline. Patient and Observations provide the context, CarePlan captures the output as recommendations and  Subscription refers to any additional systems that should receive a push notification from the system. The adoption of a standardized FHIR bundle will promote automated service discovery and use.

\section{Serverless on FHIR (architecture)}

Deploying models as small, stateless, containerized microservices ensure maintainability. The scalability of deployments is vital to ensure efficient use of resources. The serverless architecture provides ‘backend as a service’ with the cloud provider managing all backend infrastructure by automatically scaling services according to demand, from zero to infinity. The \emph{Function as a Service} with small, stateless, polyglot, containerized functions compiled for 64 bit Linux and responding to requests on port 8080, ensures portability.

Finally, the FHIR  specification can be used to declare the capabilities of service. The digital health services should respond with a FHIR EndPoint on a ‘GET’ request declaring the capabilities and metadata. Inputs and Outputs should be FHIR bundles with the Patient, Observation, CarePlan and Subscription resources. The Patient resource provides the context for personalized output, Observation resources provide input data and CarePlan encapsulates the output. The optional Subscription resource points to potential subscribing systems.

The proposed four-tier architecture for cloud-based digital health service (see Figure 1) with containerized microservices for maintainability, serverless architecture for scalability, function as a service for portability and FHIR specifications for declaring capabilities. We call this architecture \emph{Serverless on FHIR} and propose this as a standard to deploy ML and AI algorithms for digital health that can be consumed by downstream systems such as EMRs and visualization tools.

\section{Industry standards and challenges}

There is no current standard for deploying ML models in healthcare. Models are usually prepared by the analytics team using specialized hardware and software. The technical team deploys it using customized tools or proprietary standards. There may be little interaction between the two teams during this approach to development and implementation, leading to infrequent model updates. There are no standards for service discovery and models are generally deployed inhouse for a limited audience. This siloed approach leads to duplication of work even within the same organization. The progress made in model building using the latest ML tools and techniques may make hardly any impact at the point of patient care.

\emph{Open Neural Network Exchange} (ONNX) \cite{onnx2017website} is an open standard for neural network models. ONNX has a set of tools for converting models generated using other frameworks into a standard file format. Several runtimes and compilers are available for ONNX models for more efficient deployments \cite{lin2019onnc}. The ONNX standard strives to make models, framework agnostic, helping AI developers to share their models.

\emph{Docker} containers have democratized virtualization — the process of creating a virtual instance in a host \cite{bernstein2014containers}. Docker provides a set of tools for creating and using containers and it has become crucial in successful DevOps \cite{kang2016container}. Docker provides operating system-level virtualization that is more efficient than hardware level virtualization \cite{merkel2014docker}. Another advantage of Docker is the ease with which containers can be versioned and pushed to repositories for sharing. Docker has become the de-facto standard for deploying containers on the cloud.

\emph{Kubernetes} is an orchestrating framework for docker containers by decoupling containers from the infrastructure on which they run \cite{gogouvitis2018seamless}. Kubernetes facilitates the scaling of services by increasing or decreasing the number of deployed containers. Kubernetes also makes networking between containers within the host easy and exposes external services through a shared host address. OpenFaaS® \cite{ellis2019openfaas} is a set of tools that makes it easy to deploy polyglot functions and microservices to Kubernetes. It adopts a set of standards that makes microservice deployment agnostic to the underlying cloud vendor.
Stateful models --- models that persist state in a database --- may require further considerations during deployments. Databases, traditionally are not amenable to scaling down to zero. Kubernetes and similar container orchestrating platforms manage databases with persistent volumes and persistent volume claims \cite{mohamed2019extensible}. 

\subsection{SMART-ON-FHIR}

The Substitutable Medical Applications and Reusable Technologies (SMART) for platform-agnostic medical applications adopting the FHIR schema is called the SMART on FHIR platform \cite{mandel2016smart}. The SMART on FHIR utilizes a profiled version of FHIR and OAuth2 for authentication. Serverless on FHIR describes the low-level deployment architecture for microservices that can be consumed by SMART apps because of the use of a FHIR schema.

Privacy and security considerations are essential for any clinical deployment. High-level applications such as EMRs determine the privacy and security standards. SMART on FHIR adopts OAuth2 and OpenID for authorization and authentication, respectively \cite{mandel2016smart}, and the orchestrating frameworks such as Kubernetes and Docker Swarm implement them for services \cite{modak2018techniques}.

\section{Use cases and examples}

To demonstrate the use of our framework for deployment, we describe how to build and deploy a machine learning model from a national database. Discharge Abstract Database (DAD) is a Canada-wide database of hospital admission and discharge data excluding the province of Quebec, maintained by the Canadian Institute for Health Information (CIHI) \cite{joseph2009validation}. The data points in DAD include patient demographics, comorbidities coded in the International Statistical Classification of Diseases and Related Health Problems (ICD), interventions encoded in the Canadian Classification of Health Interventions (CCI) and the length of stay. CIHI provides DAD in the SPSS (.sav) format with each record having horizontal fields for 20 comorbidities and 25 interventions.

As a first step, we vectorized data with each record represented as patient demographics (age and gender), morbidities, interventions and length of stay. All variables were converted into binary (age over 40, length of stay exceeding 10 days, morbidities represented by first 3 characters of their respective ICD codes and the interventions represented by the first 3 characters of their respective CCI code).


Automatic Machine Learning (AutoML) --- the process of automating the ML model building --- is becoming increasingly popular in areas such as healthcare and is reducing the barrier of entry to ML. We used H2O.ai \cite{h2oai2019}, a popular AUTOML platform to create the model. H2O.ai can compare a range of models and display a leaderboard with various prediction metrics, that helps in choosing the appropriate model. We chose the XGBoost \cite{chen2016xgboost} method as it had maximum area under the curve (AUC) of 0.92, and used H2O.ai to build the final model. Deciding on the right model for various scenarios may require considering the other parameters, but we chose XGBoost for demonstration purposes. The model was subsequently exported to the proprietary MOJO (Model Object, Optimized) format \cite{kochura2017performance}.

We used the OpenFaaS tools and a custom template we created to build a container with the model and a wrapper for serving predictions over http on port 8080. The OpenFaaS tools allow the deployment of the prediction workflow 
on frameworks such as Kubernetes and Docker Swarm. Most serverless platforms support these standards. The process of containerization and deployment using the OpenFaaS toolset ensures maintainability, scalability and portability. When more data becomes available a different model can be deployed with minimal effort.

A gateway container that wraps incoming and outgoing messages in a FHIR bundle will ensure discoverability. FHIR is an evolving specification and separating the FHIR wrapper helps in updating FHIR wrapper without changing any of the existing containers. The OpenFaaS template for deploying the model is made available as open-source at https://github.com/dermatologist/java-ext \cite{openfaasbeapen2019} for modification and use.

\section{Discussion}

ML is becoming increasingly prevalent in healthcare. It brings to light ethical issues different from that of other domains \cite{vayena2018machine}. Advanced ML models need to be deployed in such a way that they are available for everyone. But bias in ML models towards populations that are not widely represented is a known problem \cite{adamson2018machine}. Inefficient deployments can worsen healthcare inequalities by making ML available only to a privileged few.

Our proposed framework is based on technologies that are currently available. The turnover time for technological innovations in ML is short. In the future, browser-based ML and on-device and edge computing may become more frequent, and need for cloud deployments may become less critical \cite{jeong2018computation}.

There is an increasing emphasis by healthcare providers on patient-centred solutions enabling patients to make informed decisions. ML is popular in pre-clinical research, and recent advances in drug discovery and bioinformatics rely on ML algorithms \cite{larranaga2006machine}. The time to characterize genes and develop lab tests and vaccines has been greatly reduced, as seen in the recent COVID-19 pandemic. However, there is growing discontent in the clinical community that their participation has not been encouraged \cite{zorc2019machine}. Many of the ML algorithms developed for clinical decision making are only available to a few, and efficient ML deployments may reduce this disparity.

FHIR, as an interoperability solution, has gained wide acceptance, which is why we adopted FHIR as the backbone of our deployment framework. We have extended BI  deployment guidelines \cite{8125559} to eHealth but have refrained from offering any new standard. We hope that the \emph{Serverless on FHIR} that we propose will evolve and mature into a standard, with a focus on serving clinicians.


ML and AI have resulted in several innovative applications in medicine. But many of these innovations need support to scale from computer labs to the bedside. Traditionally, deployment of ML models has received less attention than model building. We have proposed a taxonomy of ML models and a set of deployment principles to bridge the gap between researchers who build innovative healthcare models and implementation engineers who make them available at the point of patient care.

\bibliographystyle{unsrt}  
\bibliography{references}  

\begin{thebibliography}{10}

\bibitem{yu2018artificial}
Kun-Hsing Yu, Andrew~L Beam, and Isaac~S Kohane.
\newblock Artificial {I}ntelligence in healthcare.
\newblock {\em Nature Biomedical Engineering}, 2(10):719--731, 2018.

\bibitem{chen2019develop}
Po-Hsuan~Cameron Chen, Yun Liu, and Lily Peng.
\newblock How to develop machine learning models for healthcare.
\newblock {\em Nature Materials}, 18(5):410, 2019.

\bibitem{o2008gartner}
Daniel~E O'Leary.
\newblock Gartner's hype cycle and information system research issues.
\newblock {\em International Journal of Accounting Information Systems},
  9(4):240--252, 2008.

\bibitem{ben2019impact}
David Ben-Israel, W~Bradley Jacobs, Steve Casha, Stefan Lang, Won Hyung~A Ryu,
  Madeleine de~Lotbiniere-Bassett, and David~W Cadotte.
\newblock The impact of machine learning on patient care: a systematic review.
\newblock {\em Artificial Intelligence in Medicine}, page 101785, 2019.

\bibitem{tang2019recent}
Binhua Tang, Zixiang Pan, Kang Yin, and Asif Khateeb.
\newblock Recent advances of deep learning in bioinformatics and computational
  biology.
\newblock {\em Frontiers in Genetics}, 10:214, 2019.

\bibitem{shen2017deep}
Dinggang Shen, Guorong Wu, and Heung-Il Suk.
\newblock Deep learning in medical image analysis.
\newblock {\em Annual Review of Biomedical Engineering}, 19:221--248, 2017.

\bibitem{kwakernaak2020using}
Sascha Kwakernaak, Kasper van Mens, Wiepke Cahn, and Richard Janssen.
\newblock Using machine learning to predict mental healthcare consumption in
  non-affective psychosis.
\newblock {\em Schizophrenia Research}, S09209964(20):30022--23, 2020.

\bibitem{sboner2003multiple}
Andrea Sboner, Claudio Eccher, Enrico Blanzieri, Paolo Bauer, Mario
  Cristofolini, Giuseppe Zumiani, and Stefano Forti.
\newblock A multiple classifier system for early melanoma diagnosis.
\newblock {\em {Artificial Intelligence in Medicine}}, 27(1):29--44, 2003.

\bibitem{kourou2015machine}
Konstantina Kourou, Themis~P Exarchos, Konstantinos~P Exarchos, Michalis~V
  Karamouzis, and Dimitrios~I Fotiadis.
\newblock Machine learning applications in cancer prognosis and prediction.
\newblock {\em Computational and Structural Biotechnology Journal}, 13:8--17,
  2015.

\bibitem{weng2002protein}
Zhiping Weng and Charles DeLisi.
\newblock Protein therapeutics: promises and challenges for the 21st century.
\newblock {\em Trends in Biotechnology}, 20(1):29--35, 2002.

\bibitem{eapen2020public}
Bell~Raj Eapen, Norm Archer, and Kamran Sartipi.
\newblock Public health information systems: From data to knowledge.
\newblock {\em Macsphere: McMaster University’s Institutional Repository:
  11375/25272}, 2020.

\bibitem{wang2012machine}
Shijun Wang and Ronald~M Summers.
\newblock Machine learning and radiology.
\newblock {\em Medical Image Analysis}, 16(5):933--951, 2012.

\bibitem{roopa2017survey}
CK~Roopa and BS~Harish.
\newblock A survey on various machine learning approaches for ecg analysis.
\newblock {\em International Journal of Computer Applications}, 163(9):25--33,
  2017.

\bibitem{muller2008machine}
Klaus-Robert M{\"u}ller, Michael Tangermann, Guido Dornhege, Matthias
  Krauledat, Gabriel Curio, and Benjamin Blankertz.
\newblock Machine learning for real-time single-trial eeg-analysis: from
  brain--computer interfacing to mental state monitoring.
\newblock {\em Journal of Neuroscience Methods}, 167(1):82--90, 2008.

\bibitem{obermeyer2016predicting}
Ziad Obermeyer and Ezekiel~J Emanuel.
\newblock Predicting the future—big data, machine learning, and clinical
  medicine.
\newblock {\em The New England Journal of Medicine}, 375(13):1216, 2016.

\bibitem{kaur2006empirical}
Harleen Kaur and Siri~Krishan Wasan.
\newblock Empirical study on applications of data mining techniques in
  healthcare.
\newblock {\em Journal of Computer Science}, 2(2):194--200, 2006.

\bibitem{geron2019hands}
Aur{\'e}lien G{\'e}ron.
\newblock {\em Hands-On Machine Learning with Scikit-Learn, Keras, and
  TensorFlow: Concepts, Tools, and Techniques to Build Intelligent Systems}.
\newblock O'Reilly Media, 2019.

\bibitem{kim2017development}
Seong~Jae Kim, Kyong~Jin Cho, and Sejong Oh.
\newblock Development of machine learning models for diagnosis of glaucoma.
\newblock {\em PloS one}, 12(5), 2017.

\bibitem{betancur2018prognostic}
Julian Betancur, Yuka Otaki, Manish Motwani, Mathews~B Fish, Mark Lemley,
  Damini Dey, Heidi Gransar, Balaji Tamarappoo, Guido Germano, Tali Sharir,
  et~al.
\newblock Prognostic value of combined clinical and myocardial perfusion
  imaging data using machine learning.
\newblock {\em JACC: Cardiovascular Imaging}, 11(7):1000--1009, 2018.

\bibitem{toussi2009using}
Massoud Toussi, Jean-Baptiste Lamy, Philippe Le~Toumelin, and Alain Venot.
\newblock Using data mining techniques to explore physicians' therapeutic
  decisions when clinical guidelines do not provide recommendations: methods
  and example for type 2 diabetes.
\newblock {\em BMC Medical Informatics and Decision Making}, 9(1):28, 2009.

\bibitem{nair2018applying}
Lekha~R Nair, Sujala~D Shetty, and Siddhanth~D Shetty.
\newblock Applying spark based machine learning model on streaming big data for
  health status prediction.
\newblock {\em Computers \& Electrical Engineering}, 65:393--399, 2018.

\bibitem{yarkoni2017choosing}
Tal Yarkoni and Jacob Westfall.
\newblock Choosing prediction over explanation in psychology: Lessons from
  machine learning.
\newblock {\em Perspectives on Psychological Science}, 12(6):1100--1122, 2017.

\bibitem{sahoo2020automatic}
Santanu Sahoo, Asit Subudhi, Manasa Dash, and Sukanta Sabut.
\newblock Automatic classification of cardiac arrhythmias based on hybrid
  features and decision tree algorithm.
\newblock {\em International Journal of Automation and Computing}, pages 1--11,
  2020.

\bibitem{shmueli2011predictive}
Galit Shmueli and Otto~R Koppius.
\newblock Predictive analytics in information systems research.
\newblock {\em MIS Quarterly}, pages 553--572, 2011.

\bibitem{choi2016retain}
Edward Choi, Mohammad~Taha Bahadori, Jimeng Sun, Joshua Kulas, Andy Schuetz,
  and Walter Stewart.
\newblock Retain: An interpretable predictive model for healthcare using
  reverse time attention mechanism.
\newblock In {\em Advances in Neural Information Processing Systems}, pages
  3504--3512, 2016.

\bibitem{freeman2020algorithm}
Karoline Freeman, Jacqueline Dinnes, Naomi Chuchu, Yemisi Takwoingi, Sue~E
  Bayliss, Rubeta~N Matin, Abhilash Jain, Fiona~M Walter, Hywel~C Williams, and
  Jonathan~J Deeks.
\newblock Algorithm based smartphone apps to assess risk of skin cancer in
  adults: systematic review of diagnostic accuracy studies.
\newblock {\em BMJ}, 368, 2020.

\bibitem{hekler2019deep}
Achim Hekler, Jochen~S Utikal, Alexander~H Enk, Wiebke Solass, Max Schmitt,
  Joachim Klode, Dirk Schadendorf, Wiebke Sondermann, Cindy Franklin, Felix
  Bestvater, et~al.
\newblock Deep learning outperformed 11 pathologists in the classification of
  histopathological melanoma images.
\newblock {\em European Journal of Cancer}, 118:91--96, 2019.

\bibitem{varshney2017safety}
Kush~R Varshney and Homa Alemzadeh.
\newblock On the safety of machine learning: Cyber-physical systems, decision
  sciences, and data products.
\newblock {\em Big Data}, 5(3):246--255, 2017.

\bibitem{sutton2011reinforcement}
Richard~S Sutton and Andrew~G Barto.
\newblock {\em Reinforcement Learning: An introduction}.
\newblock Cambridge, MA: MIT Press, 2011.

\bibitem{saria2018individualized}
Suchi Saria.
\newblock Individualized sepsis treatment using reinforcement learning.
\newblock {\em Nature Medicine}, 24(11):1641--1642, 2018.

\bibitem{csaky2019deep}
Richard Csaky.
\newblock Deep learning based chatbot models.
\newblock {\em arXiv preprint arXiv:1908.08835}, 2019.

\bibitem{roca2020microservice}
Surya Roca, Jorge Sancho, Jos{\'e} Garc{\'\i}a, and {\'A}lvaro Alesanco.
\newblock Microservice chatbot architecture for chronic patient support.
\newblock {\em Journal of Biomedical Informatics}, 102:103305, 2020.

\bibitem{wallace2012deploying}
Byron~C Wallace, Kevin Small, Carla~E Brodley, Joseph Lau, and Thomas~A
  Trikalinos.
\newblock Deploying an interactive machine learning system in an evidence-based
  practice center: abstrackr.
\newblock In {\em proceedings of the 2nd ACM SIGHIT International Health
  Informatics Symposium}, pages 819--824, 2012.

\bibitem{amershi2019software}
Saleema Amershi, Andrew Begel, Christian Bird, Robert DeLine, Harald Gall, Ece
  Kamar, Nachiappan Nagappan, Besmira Nushi, and Thomas Zimmermann.
\newblock Software engineering for machine learning: A case study.
\newblock In {\em 2019 IEEE/ACM 41st International Conference on Software
  Engineering: Software Engineering in Practice (ICSE-SEIP)}, pages 291--300.
  IEEE, 2019.

\bibitem{malhotra2016software}
Ruchika Malhotra and Anuradha Chug.
\newblock Software maintainability: Systematic literature review and current
  trends.
\newblock {\em International Journal of Software Engineering and Knowledge
  Engineering}, 26(08):1221--1253, 2016.

\bibitem{dragoni2017microservices}
Nicola Dragoni, Saverio Giallorenzo, Alberto~Lluch Lafuente, Manuel Mazzara,
  Fabrizio Montesi, Ruslan Mustafin, and Larisa Safina.
\newblock Microservices: yesterday, today, and tomorrow.
\newblock In {\em Present and Ulterior Software Engineering}, pages 195--216.
  Springer, 2017.

\bibitem{merkel2014docker}
Dirk Merkel.
\newblock Docker: lightweight linux containers for consistent development and
  deployment.
\newblock {\em Linux journal}, 2014(239):2, 2014.

\bibitem{kang2016container}
Hui Kang, Michael Le, and Shu Tao.
\newblock Container and microservice driven design for cloud infrastructure
  devops.
\newblock In {\em 2016 IEEE International Conference on Cloud Engineering
  (IC2E)}, pages 202--211. IEEE, 2016.

\bibitem{ebert2016devops}
Christof Ebert, Gorka Gallardo, Josune Hernantes, and Nicolas Serrano.
\newblock Devops.
\newblock {\em IEEE Software}, 33(3):94--100, 2016.

\bibitem{bernstein2014containers}
David Bernstein.
\newblock Containers and cloud: From lxc to docker to kubernetes.
\newblock {\em IEEE Cloud Computing}, 1(3):81--84, 2014.

\bibitem{modak2018techniques}
Arsh Modak, SD~Chaudhary, PS~Paygude, and SR~Ldate.
\newblock Techniques to secure data on cloud: Docker swarm or kubernetes?
\newblock In {\em 2018 Second International Conference on Inventive
  Communication and Computational Technologies (ICICCT)}, pages 7--12. IEEE,
  2018.

\bibitem{weyuker2002metric}
Elaine~J Weyuker and Alberto Avritzer.
\newblock A metric to predict software scalability.
\newblock In {\em Proceedings Eighth IEEE Symposium on Software Metrics}, pages
  152--158. IEEE, 2002.

\bibitem{baldini2017serverless}
Ioana Baldini, Paul Castro, Kerry Chang, Perry Cheng, Stephen Fink, Vatche
  Ishakian, Nick Mitchell, Vinod Muthusamy, Rodric Rabbah, Aleksander
  Slominski, et~al.
\newblock Serverless computing: Current trends and open problems.
\newblock In {\em Research Advances in Cloud Computing}, pages 1--20. Springer,
  2017.

\bibitem{manner2018cold}
Johannes Manner, Martin Endre{\ss}, Tobias Heckel, and Guido Wirtz.
\newblock Cold start influencing factors in function as a service.
\newblock In {\em 2018 IEEE/ACM International Conference on Utility and Cloud
  Computing Companion (UCC Companion)}, pages 181--188. IEEE, 2018.

\bibitem{lynn2017preliminary}
Theo Lynn, Pierangelo Rosati, Arnaud Lejeune, and Vincent Emeakaroha.
\newblock A preliminary review of enterprise serverless cloud computing
  (function-as-a-service) platforms.
\newblock In {\em 2017 IEEE International Conference on Cloud Computing
  Technology and Science (CloudCom)}, pages 162--169. IEEE, 2017.

\bibitem{kaewkasi2018docker}
Chanwit Kaewkasi.
\newblock {\em Docker for Serverless Applications: Containerize and Orchestrate
  Functions Using OpenFaas, OpenWhisk, and Fn}.
\newblock Packt Publishing Ltd, 2018.

\bibitem{ellis2019openfaas}
A~Ellis et~al.
\newblock Openfaas: Serverless functions made simple.
\newblock \url{https://www.openfaas.com/}, 2019 (accessed April 5, 2020).

\bibitem{gcr2019serverless}
Cloud Run.
\newblock Google cloud platform.
\newblock \url{https://cloud.google.com/run}, 2019 (accessed April 5, 2020).

\bibitem{eapen2019drishti}
Bell~Raj Eapen, Norm Archer, Kamran Sartipi, and Yufei Yuan.
\newblock Drishti: a sense-plan-act extension to open mhealth framework using
  fhir.
\newblock In {\em 2019 IEEE/ACM 1st International Workshop on Software
  Engineering for Healthcare (SEH)}, pages 49--52. IEEE, 2019.

\bibitem{bender2013hl7}
Duane Bender and Kamran Sartipi.
\newblock Hl7 fhir: An agile and restful approach to healthcare information
  exchange.
\newblock In {\em Proceedings of the 26th IEEE International Symposium on
  Computer-based Medical Systems}, pages 326--331. IEEE, 2013.

\bibitem{fhir2019endpoint}
FHIR.
\newblock Resource endpoint.
\newblock \url{https://www.hl7.org/fhir/endpoint.html}, 2019 (accessed April 5,
  2020).

\bibitem{onnx2017website}
ONNX.
\newblock Open neural network exchange. the open standard for machine learning
  interoperability.
\newblock \url{https://onnx.ai/}, 2017 (accessed April 5, 2020).

\bibitem{lin2019onnc}
Wei-Fen Lin, Der-Yu Tsai, Luba Tang, Cheng-Tao Hsieh, Cheng-Yi Chou, Ping-Hao
  Chang, and Luis Hsu.
\newblock Onnc: A compilation framework connecting onnx to proprietary deep
  learning accelerators.
\newblock In {\em 2019 IEEE International Conference on Artificial Intelligence
  Circuits and Systems (AICAS)}, pages 214--218. IEEE, 2019.

\bibitem{gogouvitis2018seamless}
Spyridon~V Gogouvitis, Harald Mueller, Sreenath Premnadh, Andreas Seitz, and
  Bernd Bruegge.
\newblock Seamless computing in industrial systems using container
  orchestration.
\newblock {\em Future Generation Computer Systems}, 2018.

\bibitem{mohamed2019extensible}
Mohamed Mohamed, Robert Engel, Amit Warke, Shay Berman, and Heiko Ludwig.
\newblock Extensible persistence as a service for containers.
\newblock {\em Future Generation Computer Systems}, 97:10--20, 2019.

\bibitem{mandel2016smart}
Joshua~C Mandel, David~A Kreda, Kenneth~D Mandl, Isaac~S Kohane, and Rachel~B
  Ramoni.
\newblock Smart on fhir: a standards-based, interoperable apps platform for
  electronic health records.
\newblock {\em Journal of the American Medical Informatics Association},
  23(5):899--908, 2016.

\bibitem{joseph2009validation}
KS~Joseph and J~Fahey.
\newblock Validation of perinatal data in the discharge abstract database of
  the canadian institute for health information.
\newblock {\em Chronic diseases in Canada}, 29(3):96--100, 2009.

\bibitem{h2oai2019}
H2O.ai.
\newblock Ai and ml platform.
\newblock \url{https://www.h2o.ai/}, 2020 (accessed April 5, 2020).

\bibitem{chen2016xgboost}
Tianqi Chen and Carlos Guestrin.
\newblock Xgboost: A scalable tree boosting system.
\newblock In {\em Proceedings of the 22nd ACM SIGKDD International Conference
  on Knowledge Discovery and Data Mining}, pages 785--794, 2016.

\bibitem{kochura2017performance}
Yuriy Kochura, Sergii Stirenko, Oleg Alienin, Michail Novotarskiy, and Yuri
  Gordienko.
\newblock Performance analysis of open source machine learning frameworks for
  various parameters in single-threaded and multi-threaded modes.
\newblock In {\em Conference on Computer Science and Information Technologies},
  pages 243--256. Springer, 2017.

\bibitem{openfaasbeapen2019}
Bell~Raj Eapen.
\newblock Openfaas template for serverless on fhir.
\newblock \url{https://github.com/dermatologist/java-ext}, 2019 (accessed April
  5, 2020).

\bibitem{vayena2018machine}
Effy Vayena, Alessandro Blasimme, and I~Glenn Cohen.
\newblock Machine learning in medicine: addressing ethical challenges.
\newblock {\em PLoS Medicine}, 15(11), 2018.

\bibitem{adamson2018machine}
Adewole~S Adamson and Avery Smith.
\newblock Machine learning and health care disparities in dermatology.
\newblock {\em JAMA Dermatology}, 154(11):1247--1248, 2018.

\bibitem{jeong2018computation}
Hyuk-Jin Jeong, InChang Jeong, Hyeon-Jae Lee, and Soo-Mook Moon.
\newblock Computation offloading for machine learning web apps in the edge
  server environment.
\newblock In {\em 2018 IEEE 38th International Conference on Distributed
  Computing Systems (ICDCS)}, pages 1492--1499. IEEE, 2018.

\bibitem{larranaga2006machine}
Pedro Larranaga, Borja Calvo, Roberto Santana, Concha Bielza, Josu Galdiano,
  Inaki Inza, Jos{\'e}~A Lozano, Rub{\'e}n Armananzas, Guzm{\'a}n Santaf{\'e},
  Aritz P{\'e}rez, et~al.
\newblock Machine learning in bioinformatics.
\newblock {\em Briefings in Bioinformatics}, 7(1):86--112, 2006.

\bibitem{zorc2019machine}
Joseph~J Zorc, James~M Chamberlain, and Lalit Bajaj.
\newblock Machine learning at the clinical bedside—the ghost in the machine.
\newblock {\em JAMA Pediatrics}, 173(7):622--624, 2019.

\bibitem{8125559}
A.~{Khan}.
\newblock Key characteristics of a container orchestration platform to enable a
  modern application.
\newblock {\em IEEE Cloud Computing}, 4(5):42--48, 2017.

\end{thebibliography}

\end{document}